\documentclass[12pt]{article}
\pdfoutput=1

\usepackage{amsmath}
\usepackage{amsfonts}
\usepackage{amscd}
\usepackage{amsthm}
\usepackage{setspace}
\usepackage{amssymb}

\usepackage{graphicx}
\usepackage{authblk}
\usepackage{caption}
\usepackage{ytableau}
\usepackage{mathtools}

\begin{document}

\begin{titlepage}

\begin{flushright}

\end{flushright}

\vskip 1cm

\begin{center}

{\bf \Large Eight-dimensional non-geometric heterotic strings\\ and enhanced gauge groups}

\vskip 1.2cm

Yusuke Kimura$^1$ 
\vskip 0.4cm
{\it $^1$Tokyo Institute of Technology, International Research Frontiers Initiative, \\ Nagatsuta-cho, Midori-ku, Yokohama 226-8503, Japan}

(Dated: May 30, 2022) 

\vskip 1.5cm
\abstract{We review the construction of eight-dimensional (8D) non-geometric heterotic strings, proposed by Malmendier and Morrison, which do not allow for a geometric interpretation. In the construction, the $\mathfrak{e}_8\oplus \mathfrak{e}_7$ gauge algebra is unbroken. The moduli space of 8D non-geometric heterotic strings and theories arising in the moduli space can be analyzed by studying the geometries of elliptically fibered K3 surfaces with a global section by applying F-theory/heterotic duality. Additionally, we review the results of the points in the 8D non-geometric heterotic moduli with the unbroken $\mathfrak{e}_8\oplus \mathfrak{e}_7$ gauge algebra, at which the non-Abelian gauge groups are maximally enhanced. At these points, the gauge groups formed in the theories do not allow for a perturbative interpretation of the heterotic perspective. However, from the dual F-theory perspective, the K3 geometries at these points are deformations of the stable degenerations that arise from the coincident 7-branes. On the heterotic side, these enhancements can be understood as a non-perturbative effect of 5-brane insertions.
}

\end{center}
\end{titlepage}

\tableofcontents
\section{Introduction}
\par In this study, we discuss the gauge group in elementary particle physics. Elementary particles are fundamental subatomic particles that cannot be subdivided into smaller particles. The elementary particle theory is employed to investigate the interaction among such fundamental particles. The notion of the gauge group arises in the theoretical framework of elementary particles to describe the intersections among the particles. Studies on elementary particle physics show that gauge groups are crucial in describing the forces or interactions among elementary particles.
\par First, we introduce the notion of a group in mathematics; it refers to a set of elements such that one can ``multiply'' any two elements. This operation is associative, implying that $(g\cdot h)\cdot k=g\cdot (h\cdot k) $. A distinguished element exists in the group and is referred to as an {\it identity}, $e$, such that for any element $g$ in a group, the multiplication of $g$ by identity $e$, $g\cdot e$, is $g$: $g\cdot e=e\cdot g=g$. Furthermore, every element $g$ in the group has an inverse $g^{-1}$ such that the multiplication of $g$ with its inverse $g^{-1}$ yields identity $e$: $g\cdot g^{-1}=g^{-1}\cdot g=e$. These rules define a group. Although the description of the rules may appear slightly abstract at first, the rules are very general. Owing to this generality, many objects in mathematics are groups. Several representative examples of groups are found to be familiar notions. For example, a set of integers, $\mathbb{Z}$, with addition + regarded as ``multiplication,'' is a group. All possible permutations of $n$ objects is another example of a group, which is referred to as the symmetric group, $S_n$. Some geometric symmetries also form a group. 
\par Groups can be categorized into two classes: commutative groups and non-commutative ones. For group $G$, if the multiplication of any two elements commutes, that is, $a\cdot b=b\cdot a$ for any two elements $a$ and $b$ in $G$, then group $G$ is said to be commutative. A commutative group is also referred to as the Abelian group. $\mathbb{Z}$ (with addition) is an example of the Abelian group. When there is a pair of elements, $a$ and $b$, that does not commute, $a\cdot b\ne b\cdot a$, group $G$ is said to be non-commutative. In elementary particle physics, when gauge group $G$ is non-commutative, $G$ is referred to as a non-Abelian gauge group.  
\par Some recent developments in elementary particle physics can be attributed to string theory. This theory is considered as a candidate for the consistent quantum gravitational theory. The quantum field theoretical formulation of gravity originally had a divergence problem. To overcome this problem, the notion of point particles was replaced with that of particles with a one-dimensional structure, or ``strings,'' leading to the formulation and construction of string theory. However, string theory research has revealed that such ``string'' particles can only consistently exist in high-dimensional space-time, whose dimensions are larger than four. 
\par Because the real world is four dimensional, comprising three spatial dimensions and one time dimension, many researchers in string theory consider that the extra-dimensional space, referred to as a {\it compactification space}, is tiny and escapes experimental observations. Studies on string theory show that the geometry of the compactification space is closely related to the structure of the gauge group.
\par In this study, we aim to introduce and review some recent developments on this aspect of string theory, including our own results. 

\par String theory has several different versions of formulations. These are considered equivalent, at least under certain circumstances, and the different versions of string theories are closely related. The several formulations of string theory are not completely independent theories; instead, they can be viewed as different viewpoints of an identical theory under some conditions. The formulations of sting theory include {\it heterotic strings} \cite{Gross1984, Gross1985hetero1, Gross1985hetero2} and {\it F-theory} \cite{Vaf, MV1, MV2}. 

\par How does the geometry of the compactification space determine the gauge group structure? Based on F-theory, this point can be explained as follows: in type IIB superstring theory, Ramond-Ramond 0-form $C_0$ and dilation $\phi$ arise. Type IIB superstring theory is considered to possess $SL_2(\mathbb{Z})$ symmetry, wherein the $SL_2(\mathbb{Z})$ group acts on bosons, including these particles. In the formulation of F-theory \cite{Vaf, MV1, MV2}, Ramond-Ramond scalar $C_0$ and dilaton $\phi$ are treated in a unified manner as the axio-dilaton, $C_0+i e^{-\phi}$, which is identified with the modular parameter, $\tau$, of an elliptic curve. Because the modular parameter of an elliptic curve has $SL_2(\mathbb{Z})$ symmetry, this identification is useful for technical reasons. Here, an elliptic curve is in real dimensions a two-dimensional surface with one hole. Elliptic curves are frequently considered in algebraic geometry, wherein the geometries are considered over complex numbers in many situations. Elliptic curves have a complex dimension of 1, and owing to this property, they are referred to as curves, although they are surfaces in real dimensions. 
\par Given these backgrounds, compactification spaces in F-theory are ``elliptic fibrations,'' which mean spaces consisting of elliptic curves that lie over the base spaces. 
\par Physical objects known as ``branes'' appear in string theory. String particles can be of two types: an open string and a closed string. Open strings mediate the forces of the gauge groups. An open string has two ends on a brane. In F-theory, there are 7-branes and they are essentially related to the non-Abelian gauge group formed in the theory. 
\par 7-branes are wrapped on a subspace in the base space. Elliptic curves are generically smooth over the base space, but over a complex codimension one subspace they degenerate to singular fibers. 7-branes are wrapped on the components in this subspace of the base space.
\par The different ways in which the elliptic fibers degenerate to the singular fibers were classified for elliptically fibered complex surfaces \cite{Kod1, Kod2}. A correspondence \cite{MV2, BIKMSV} exists between the ways in which the elliptic fibers degenerate and the non-Abelian gauge groups formed on the 7-branes in F-theory. The non-Abelian gauge groups are given as Lie groups and, hence, have $ADE$ types. Kodaira's classification \cite{Kod1, Kod2} of the singular fibers of an elliptic fibration also has $ADE$ types. Non-Abelian gauge group arising on 7-branes that are wrapped on a component in the base space, and the degeneration of an elliptic fiber lying over that component, have identical $ADE$ types. 

\par Heterotic strings and F-theory are closely related through the duality relation \cite{Vaf, MV1, MV2, Sen, FMW}. Heterotic strings on the 2-torus, $T^2$, are dual to F-theory on elliptically fibered K3 surfaces. In \cite{MM}, a class of heterotic strings was constructed on $T^2$ that did not have a geometric interpretation \footnote{Discussion of connections $O^+(\Lambda^{2,2})$-modular forms and non-geometric heterotic strings can be found in \cite{MMS}. Non-geometric type II theories were studied in \cite{HMW, Plauschinn2018}.}. The authors in \cite{MM} constructed ``non-geometric'' heterotic strings with unbroken $\mathfrak{e}_8\oplus \mathfrak{e}_7$ gauge algebra and F-theory duals. The main subjects of this study are non-geometric heterotic strings \cite{MM} and the structure of heterotic/F-theory duality \cite{Vaf, MV1, MV2, Sen, FMW}. In this study, we review the construction of non-geometric heterotic strings in \cite{MM} and present some developments on this topic. The points in the eight-dimensional (8D) non-geometric heterotic moduli with unbroken $\mathfrak{e}_8\oplus \mathfrak{e}_7$ gauge algebra, at which the non-Abelian gauge groups are maximally enhanced, were determined in \cite{Kimura1810}. From the heterotic viewpoint, the gauge groups formed at these point do not allow for a perturbative interpretation \cite{Kimura1810}. The K3 surfaces at these points on the dual F-theory are deformations of stable degenerations \footnote{An elliptic K3 surface degenerates into two 1/2 K3 surfaces intersecting along an elliptic curve in the stable degeneration limit. The duality relation of heterotic strings and F-theory becomes rigorous when the stable degeneration limit is considered on the F-theory side.} \cite{FMW, AM}, which arise owing to the coincidence of 7-branes, as demonstrated in \cite{Kimura1810}. This effect can be viewed on the dual heterotic side owing to the insertion of 5-branes \cite{Kimura1810}. We also discuss these issues. Recent progress in non-geometric heterotic strings can also be found, for example, in \cite{LMV1508, FGLMM1603, MS1609, GLMM1611, FM1708, Clingher2018, Kimura201902, Clingher2020}. Related discussions on the lattice structures in K3 surfaces and the duality of heterotic strings and F-theory can be found in recent studies, e.g. in \cite{Clingher2009, Clingher2111}. Some elliptically fibered K3 surfaces with extended gauge groups also appear in the Swampland Program \cite{Bedroya2021}.

\par In section \ref{sec2}, we briefly review the construction of 8D non-geometric heterotic strings given in \cite{MM}. In sections \ref{sec3.1} and \ref{sec3.2}, we then review maximally enhanced non-Abelian gauge algebras in the 8D non-geometric heterotic strings with unbroken $\mathfrak{e}_8\oplus \mathfrak{e}_7$ gauge algebra deduced in \cite{Kimura1810}. We mention the discussion of nonperturbative effect of inserting 5-branes considered in \cite{Kimura1810}. Application of the Shioda--Tate formula \cite{Shiodamodular, Tate1, Tate2} is mentioned in section \ref{sec3.1}. Details of this argument in the context of F-theory can be found, e.g. in \cite{Kimura1802}. The global structure of the gauge groups formed in F-theory \cite{AspinwallGross, AMrational, MMTW} are also mentioned in section \ref{sec3.2}. Recent studies on this aspect with non-trivial Mordell--Weil torsions can be found, e.g. in \cite{Kimura1603, Kdisc, Morrison2021, Kimura2022}. Two potential directions for extending \cite{Kimura1810} are proposed in section \ref{sec3.3}. For the first direction of extension, a related method was discussed in \cite{KimuraMizoguchi}. The second direction involves lattice theoretic techniques. Discussions of some lattice theoretic techniques in the context of F-theory/heterotic duality can be found in \cite{KRES, Kimura2021}. We state our concluding remarks in section \ref{sec4}.

\section{Review of 8D non-geometric heterotic constructions with unbroken $\mathfrak{e}_8\oplus\mathfrak{e}_7$ gauge algebra}
\label{sec2}

\subsection{Review of non-geometric heterotic constructions by Malmendier and Morrison}
\label{sec2.1}
\par There is a duality relation between F-theory and heterotic strings \cite{Vaf, MV1, MV2, Sen, FMW}, which states that F-theory on a K3 fibered Calabi--Yau (CY) $n+1$-fold and heterotic theory on an elliptically fibered CY $n$-fold are physically equivalent. 
\par Eight-dimensional heterotic string on a 2-torus, $T^2$, and F-theory on an elliptically fibered K3 surface describe physically equivalent theories. A comparison of the moduli spaces of the two 8D theories reveals some aspects of the duality relation. 
\par The moduli space of the 8D heterotic strings is the Narain space \cite{Narain}, which is quotient
\begin{equation}
\label{Narain in 2.1}
D_{2,18}/O(\Lambda^{2,18}).
\end{equation}
Here, $D_{2,18}$ represents the symmetric space of $O(2,18)$, namely $D_{2,18}=O(2) \times O(18) \backslash O(2,18)$. The 8D F-theory moduli on elliptic K3 surfaces with a global section is given as
\begin{equation}
\label{double cover in 2.1}
D_{2,18}/O^+(\Lambda^{2,18}),
\end{equation}
where $O^+(\Lambda^{2,18})=O(\Lambda^{2,18})\cap O^+(2,18)$. The double cover of the heterotic moduli space (\ref{Narain in 2.1}) matches with the F-theory moduli space (\ref{double cover in 2.1}), and this relation defines the duality of 8D heterotic strings and F-theory. 
\par The Narain space (\ref{Narain in 2.1}) decomposes into the product of the complex structure moduli, the moduli of the Wilson line expectation values, and K\"ahler moduli in a suitable limit \cite{NSWheterotic}. The action of group $O^+(\Lambda^{2,18})$ mixes the three moduli spaces in this decomposition. As a result, heterotic strings possessing the symmetry of $O^+(\Lambda^{2,18})$ do not allow for a geometric interpretation. Such theories are ``non-geometric.'' Malmendier and Morrison \cite{MM} built such 8D non-geometric heterotic strings on the two-torus with the symmetry group that is a subgroup in $O^+(\Lambda^{2,18})$. 
\par Utilizing the F-theory/heterotic duality, they constructed the 8D non-geometric heterotic strings as duals of F-theory with unbroken $\mathfrak{e}_8\oplus\mathfrak{e}_7$ gauge algebra. 8D F-theory with unbroken $\mathfrak{e}_8\oplus\mathfrak{e}_7$ gauge algebra is realized on elliptic K3 surfaces with $E_8E_7$ singularity. The moduli space of F-theory on elliptic K3 surfaces with $E_8E_7$ singularity with a global section is given as \cite{MM}
\begin{equation}
D_{2,3}/O^+(L^{2,3}).
\end{equation}
$L^{2,3}$ denotes the orthogonal complement of lattice $E_8\oplus E_7$ inside $\Lambda^{2,18}$. There is an isomorphism between $D_{2,3}$ and the Siegel upper half-space of genus two, $\mathbb{H}_2$, i.e. $D_{2,3}\cong \mathbb{H}_2$. This isomorphism induces the correspondence of the ring of $O^+(L^{2,3})$-modular forms and the Siegel modular forms with genus two of even weight \cite{Vinberg}. 
\par An elliptically fibered K3 surface with a global section can be described by using the Weierstrass equation, and those with the $E_8E_7$ singularity are given by the following Weierstrass form:
\begin{equation}
\label{Weierstrass in 2.1}
y^2=x^3+(\alpha t^4+\gamma t^3) x+t^7+\beta t^6+\delta t^5.
\end{equation}
The specified values for $\alpha$, $\beta$, $\gamma$, and $\delta$ for the K3 surface determine a point in $D_{2,3}$, which via the isomorphism $D_{2,3}\cong \mathbb{H}_2$ specifies a point in $\mathbb{H}_2$. We denote the specified point in $\mathbb{H}_2$ by $\underline{\tau}$ to align it with the notation used in \cite{MM}. For this, $\alpha$, $\beta$, $\gamma$, and $\delta$ are given in terms of Igusa's generators \cite{Igusa} as follows \cite{Kumar, ClingherDoran2011, ClingherDoran2012} up to scale factors:
\begin{equation}
\label{section Siegel in 2.1}
\alpha=\frac{-\psi_4(\underline{\tau})}{48}, \hspace{4.8mm} \beta=\frac{-\psi_6(\underline{\tau})}{864}, \hspace{4.8mm} \gamma=-4\chi_{10}(\underline{\tau}), \hspace{4.8mm} \delta=\chi_{12}(\underline{\tau}).
\end{equation}
\par One then considers a line bundle, $\mathcal{L}$, on a two-torus $T^2$, and select sections, $\alpha, \beta, \gamma, \delta$, of $\mathcal{L}^{\otimes 4}, \mathcal{L}^{\otimes 6}, \mathcal{L}^{\otimes 10}, \mathcal{L}^{\otimes 12}$, respectively. Compactification on $T^2$ with the identification of the sections (\ref{section Siegel in 2.1}) yields the construction of non-geometric heterotic strings with $O^+(L^{2,3})$-symmetry given in \cite{MM}. 

\subsection{5-branes and enhancements in non-Abelian gauge algebra}
\label{sec2.2}
When the defining equation of an elliptic K3 surface is given, by computing the discriminant of the equation one can determine the singularity type of the surface. When the Weierstrass equation of the surface, $y^2=x^3+fx+g$, is given, then the discriminant is described by
\begin{equation}
\Delta = 4f^3+27g^2. 
\end{equation}
This particularly means that the discriminant of the K3 surface (\ref{Weierstrass in 2.1}) is given by
\begin{equation}
\label{discriminant in 2.2}
\Delta=t^9 \big(4(\alpha t+\gamma)^3 + 27t (t^2+\beta t+\delta)^2\big).
\end{equation}
For generic values of $\alpha, \beta, \gamma, \delta$, i.e. for generic points in the complex structure moduli space, the singularity type is $E_8E_7$. For special values of $\alpha, \beta, \gamma, \delta$ (equivalently, at the corresponding specific points in the moduli,) the singularity types are enhanced. This physically means that the gauge algebras are enhanced at those points. 
\par From the dual heterotic perspective, the enhanced gauge algebra can be attributed to the presence of 5-branes \cite{MM}. From the discriminant (\ref{discriminant in 2.2}), the authors of \cite{MM} deduced the 5-brane solutions in the moduli space. The two types of 5-brane solutions deduced in \cite{MM} are as follows:
\begin{eqnarray}
\label{5-branes in 2.2}
\gamma & =0 \\ \nonumber
q & =0.
\end{eqnarray}
Here, $q$ denotes a polynomial in coefficients $\alpha, \beta, \gamma$, and $\delta$ in the Weierstrass equation (\ref{Weierstrass in 2.1}), given as follows:
\begin{equation}
\label{def q in 2.2}
\begin{aligned}
q  = & 11664\, \delta^5 -5832\, \beta^2\delta^4 +864\, \alpha^3\delta^4+16\, \alpha^6\delta^3 \\
 & +216\, \alpha^3\beta^2\delta^3+16200\, \alpha\gamma^2\delta^3-2592\, \alpha^2\beta\gamma\delta^3+729\, \beta^4\delta^3 \\
 & -5670\, \alpha\beta^2\gamma^2\delta^2-13500\, \beta\gamma^3\delta^2+216\, \alpha^4\beta^2\gamma^2\delta+2700\, \alpha^2\beta^2\gamma^4-3420\, \alpha^3\beta\gamma^3\delta \\
 & +4125\, \alpha^2\gamma^4\delta+729\, \alpha\beta^4\gamma^2\delta+888\, \alpha^4\gamma^2\delta^2+6075\, \beta^3\gamma^3\delta-16\, \alpha^6\beta\gamma^3+16\, \alpha^5\gamma^4\\
 & -216\, \alpha^3\beta^3\gamma^3+16\, \alpha^7\gamma^2\delta-5625\, \alpha\beta\gamma^5-729\, \beta^5\gamma^3+3125\, \gamma^6.
\end{aligned}
\end{equation}

\section{Maximally enhanced non-Abelian gauge groups in 8D non-geometric heterotic strings with unbroken $\mathfrak{e}_8\oplus \mathfrak{e}_7$}
\label{sec3}
\subsection{Points of most enhanced non-Abelian gauge algebras in the 8D non-geometric moduli space and F-theory duals}
\label{sec3.1}
Herein, we discuss the points in the 8D F-theory moduli with unbroken $\mathfrak{e}_8\oplus \mathfrak{e}_7$ gauge algebra constructed in \cite{MM}, wherein the non-Abelian gauge algebras are maximally enhanced. These points in the moduli are special, that is, the gauge groups for the dual heterotic theories at the points do not allow for the perturbative interpretations \cite{Kimura1810}. We briefly discuss this point. 
\par First, we explain a method \cite{Kimura1810} for determining the points in the moduli at which the non-Abelian gauge algebras are maximally enhanced. The maximally enhanced non-Abelian gauge algebras are also deduced by applying the method mentioned in \cite{Kimura1810}. To achieve this goal, a modern technique in algebraic geometry was used. As reviewed in section \ref{sec2.1}, Siegel modular forms played a role in the construction in \cite{MM} to investigate the aspects of the heterotic/F-theory duality. The classification of the elliptic fibrations of (some of) K3 surfaces with Picard number 20 is another useful mathematical technique that clarifies the duality relation of F-theory and heterotic strings. 
\par Every K3 surface, $S$, has a Picard number, $\rho(S)$, which varies over the complex structure moduli. The set of K3 surfaces with a higher Picard number has a lower dimension in the complex structure moduli. Quantitatively, the dimension of the complex structure moduli of the complex K3 surfaces with Picard number $\rho$ is $20-\rho$. Intuitively, a K3 surface with a higher Picard number is rarer in terms of the whole complex structure moduli. 
\par The highest Picard number for complex K3 surfaces is 20, and they constitute a discrete set of points in the complex structure moduli. 
\par Physically, the Picard number of an elliptic K3 surface is directly related to the gauge group formed in F-theory on that K3 surface. There is an equality known as the Shioda--Tate formula \cite{Shiodamodular, Tate1, Tate2} for an elliptic K3 surface with a global section, which holds between the Picard number, rank of lattice generated by the components in the singular fibers not meeting the zero section, and the rank of the Mordell--Weil group. A global section of an elliptic K3 surface is intuitively a copy of the base $\mathbb{P}^1$, and when an elliptic fibration admits a global section, one can define an ``addition'' of any two global sections. The set of global sections under the ``addition'' form a group, referred to as the Mordell--Weil group. The rank of the Mordell--Weil group of an elliptic K3 surface yields the number of U(1) factors formed in F-theory on that K3 \cite{MV2}. Physically, the Shioda--Tate formula implies that for F-theory on an elliptic K3 surface $S$, the sum of the rank of the non-Abelian gauge group and the number of U(1)s coming from the Mordell--Weil group is equal to the Picard number $\rho(S)$ minus 2. 
\par This indicates that the sum of the ranks of the non-Abelian gauge groups and U(1)s increases as the Picard number of K3 surface increases, with the maximum of the sum being $20-2=18$. 
\par The result in mathematics \cite{PS-S, SI} showed that the complex K3 surfaces of the highest Picard number 20 are labelled by triplets of integers (modulo an appropriate group). This comes from the fact that such K3 surfaces are specified by the ``transcendental lattices'' of their own \cite{PS-S, SI}, the intersection forms of which are $2 \times 2$ integer matrices. In brief, the complex K3 surfaces of the highest Picard number 20 are labelled by the intersection forms \cite{PS-S, SI}:
\begin{equation}
\begin{pmatrix}
2a & b \\
b & 2c
\end{pmatrix},
\end{equation}
where $a,b,c$ are integers. $a\ge c\ge b\ge 0$ can be assumed by using the $GL_2(\mathbb{Z})$ action. 
\par A complex elliptic K3 surface $f: S_{\rm K3}\rightarrow \mathbb{P}^1$ of Picard number 20 with a section is referred to as {\it extremal} when the Mordell--Weil group of the fibration $f$, $MW(S_{\rm K3}, f)$, is finite. F-theory on an extremal fibration has non-Abelian gauge group of rank 18, that is the highest for 8D F-theory on an elliptic K3 surface. 
\par The complex structures of the K3 surfaces that admit extremal fibrations are completely classified in \cite{SZ}. This result can be used to determine the points in the 8D F-theory moduli with unbroken $\mathfrak{e}_8\oplus \mathfrak{e}_7$ gauge algebra, wherein the non-Abelian gauge algebras are maximally enhanced. In \cite{Kimura1810}, the following were deduced:
\begin{equation}
\mathfrak{e}_8\oplus \mathfrak{e}_7\oplus \mathfrak{su}(3)\oplus \mathfrak{su}(2), \hspace{5mm} \mathfrak{e}_8\oplus \mathfrak{e}_7\oplus \mathfrak{su}(4), \hspace{5mm} \mathfrak{e}_8^2\oplus \mathfrak{su}(3), \hspace{5mm} \mathfrak{e}_8^2\oplus \mathfrak{su}(2)^2. 
\end{equation}
\par Considering the rank of the gauge groups at these points, we can say that the gauge groups on the dual heterotic strings do not allow for a perturbative interpretation \cite{Kimura1810}. As demonstrated in \cite{Kimura1810}, the K3 surfaces at these points on the dual F-theory are deformations of the stable degenerations that arise owing to the coincidence of the 7-branes. This can be attributed to the non-perturbative effects of the insertion of 5-branes from the dual heterotic perspective \cite{Kimura1810}. 

\subsection{Sample model}
\label{sec3.2}
We describe the point in the moduli space with unbroken $\mathfrak{e}_8\oplus \mathfrak{e}_7$ gauge algebra, where the 8D F-theory has $\mathfrak{e}_8\oplus \mathfrak{e}_7\oplus \mathfrak{su}(3)\oplus \mathfrak{su}(2)$ gauge algebra as a sample. This model was studied in \cite{Kimura1810}. Based on the condition that an elliptic K3 surface with a section yielding this model has the corresponding singularity type $E_8E_7A_2A_1$, the Weierstrass equation of the K3 surface was deduced in \cite{Kimura1810} as follows:
\begin{equation}
\label{Weierstrass 602 in 3.2}
y^2=x^3+\sqrt[3]{4}t^3(\frac{-15t}{4}+3)x+t^5 (t^2+\frac{-5t}{2}+10)
\end{equation}
with discriminant 
\begin{equation}
\label{discriminant 602 in 3.2}
\Delta \sim t^9(t-4)^2(t+1)^3.
\end{equation}
In \cite{Kimura1810}, the transcendental lattice of the extremal K3 fibration (\ref{Weierstrass 602 in 3.2}) was deduced to have the following intersection form: $\begin{pmatrix}
6 & 0 \\
0 & 2
\end{pmatrix}$. Equations (\ref{Weierstrass 602 in 3.2}) and (\ref{discriminant 602 in 3.2}) imply \cite{Kimura1810} that the 7-brane stack supporting the $\mathfrak{e}_8$ algebra is located at $t=\infty$ in base $\mathbb{P}^1$; the 7-brane stack supporting the $\mathfrak{e}_7$ algebra is at $t=0$; the 7-brane stack supporting the $\mathfrak{su}(3)$ algebra is at $t=-1$; the 7-brane stack supporting the $\mathfrak{su}(2)$ algebra is at $t=4$. In \cite{Kimura1810}, the Weierstrass equation (\ref{Weierstrass 602 in 3.2}) was confirmed to satisfy a 5-brane solution \cite{MM} $q=0$ in (\ref{5-branes in 2.2}). 
\par To determine the global structure \cite{AMrational} of the gauge group formed in F-theory, one needs to specify the torsion part of the Mordell--Weil group of an elliptic fibration on which the theory is compactified. In fact, for the situation where the BPS solitons form a faithful representation of the gauge group $G$, then for F-theory on an elliptic fibration the fundamental group, $\pi_1(G)$, of the gauge group is isomorphic to the Mordell--Weil torsion of that elliptic fibration, as discussed in \cite{AMrational}. The Mordell--Weil torsion of the K3 surface with the singularity type $E_8E_7A_2A_1$ is trivial \cite{SZ}. Therefore, the global structure of the gauge group formed in F-theory on the K3 surface (\ref{Weierstrass 602 in 3.2}) is \cite{Kimura1810} $E_8\times E_7\times SU(3)\times SU(2)$.
\par As demonstrated in \cite{Kimura1810}, the elliptic K3 surface (\ref{Weierstrass 602 in 3.2}) arises as a deformation of sum $X_{[II^*, 1, 1]}\cup X_{[III^*, 2, 1]}$, wherein a 7-brane over which type $I_1$ fiber lies and 7-brane over which type $I_2$ fiber lies coincide, and two 7-branes over which type $I_1$ fibers lie become coincident. Here, $X_{[II^*, 1, 1]}$ and $X_{[III^*, 2, 1]}$ represent a rational elliptic surface with one $II^*$ type fiber and two $I_1$ type fibers \cite{MP}, and a rational elliptic surface with one $III^*$ type fiber, one $I_2$ type fiber, and one $I_1$ type fiber, respectively. This indicates that the K3 surface (\ref{Weierstrass 602 in 3.2}) arises as a deformation of stable degeneration as a result of coincident 7-branes. The effect of the coincident 7-branes and gauge enhancements on the heterotic side corresponds to the insertion of 5-branes.

\subsection{Possible extensions}
\label{sec3.3}
We discuss a few possible directions for extending the study on the enhancements in gauge groups in the 8D moduli space in \cite{Kimura1810}.
\par Investigating the maximal enhancements of gauge algebras in the 8D F-theory moduli with unbroken $\mathfrak{e}_8\oplus\mathfrak{e}_7$ gauge algebra, where the maximally enhanced gauge algebras are not purely non-Abelian, but also include U(1) factors may be interesting. 
\par Such points in the moduli space can be found by constructing a moduli space of elliptically fibered K3 surfaces with a section, whose singularity types include $E_8E_7$ with rank 17. A K3 surface at a generic point of this moduli space has Picard number 19, and the Mordell--Weil rank is zero. Subsequently, one searches for special points in the moduli space, where the corresponding K3 surfaces acquire additional independent global sections. The Mordell--Weil ranks of the K3 surface increase to one at these special points in the moduli space, and U(1) forms in F-theory on those points. This yields the construction of F-theory models with unbroken $\mathfrak{e}_8\oplus\mathfrak{e}_7$ gauge algebra, wherein the gauge groups are maximally enhanced with one U(1) factor. 
\par For example, one can construct the moduli space of elliptic K3 surfaces with a global section with singularity type $E_8E_7A_2$, and one determines points in this moduli where the Mordell--Weil rank rises to one. $\mathfrak{e}_8\oplus\mathfrak{e}_7\oplus \mathfrak{su}(3)\oplus\mathfrak{u}(1)$ gauge algebra forms at those points in the moduli. 

\par Another potential direction is to analyze points in the moduli where the gauge algebras are enhanced to rank 17. The lattice-theoretical approach can be used in this direction to help study models with gauge groups of rank 17. 
\par The lattice called the {\it N\'eron--Severi lattice} plays a role in this approach. The N\'eron--Severi lattice is a sublattice of the K3 lattice $\Lambda_{\rm K3}$, where the K3 lattice $\Lambda_{\rm K3}$ is the second integral cohomology group $H^2(K3, \mathbb{Z})$ with the structure \cite{Mil}
\begin{equation}
\Lambda_{\rm K3} \cong E_8(-1)^2 \oplus \begin{pmatrix}
0 & 1 \\
1 & 0 
\end{pmatrix}^3. 
\end{equation}
\par The N\'eron--Severi lattice, $NS(S)$, contains a considerable amount of geometric information of a K3 surface, $S$. When there is a lattice embedding of $H\oplus G$ inside the N\'eron--Severi lattice, $NS(S)$, where $H$ is the hyperbolic plane and the image of $H$ inside $NS(S)$ contains a pseudo-ample class \cite{MM}, the lattice embedding ensures \cite{PS-S} that the K3 surface has an elliptic fibration with a section of singularity type $G$. 
\par Therefore, one can determine whether the $\mathfrak{g}$ gauge algebra forms in F-theory on an elliptic K3 surface with a section by analyzing whether the N\'eron--Severi lattice of that K3 surface contains $H\oplus G$ ($H$ in the N\'eron--Severi lattice must include a pseudo-ample class), where $G$ is the corresponding singularity type of the $\mathfrak{g}$ gauge algebra.
\par According to the classification result of the singularity types of the elliptic K3 surfaces in \cite{Shimada}, the K3 surfaces in the dual 8D F-theory moduli with unbroken $\mathfrak{e}_8\oplus \mathfrak{e}_7$ gauge algebra with enhanced non-Abelian gauge groups of rank 17 must have one of the following three singularity types: $E_8^2A_1$, $E_8E_7A_1^2$, and $E_8E_7A_2$. The heterotic duals of F-theory on K3 surfaces with a section possessing these singularity types should satisfy at least one of the two 5-brane solutions (\ref{5-branes in 2.2}). 
\par Elliptic K3 surfaces with a section with each of these singularity types constitute a one-dimensional (1D) complex structure moduli. At special points in these 1D moduli spaces, either the singularity type is enhanced, resulting in an enhancement in gauge algebra, or the Mordell--Weil rank increases to one to form U(1).

\section{Concluding remarks}
\label{sec4}
In this study, we reviewed the construction of 8D non-geometric heterotic strings and dual F-theory with unbroken $\mathfrak{e}_8\oplus\mathfrak{e}_7$ gauge algebra and certain relevant studies. We also discussed related concepts and potential directions for future extensions. We believe that this article clarifies that previous studies have made progress in elucidating some aspects of heterotic/F-theory duality. 
\par The 8D non-geometric heterotic strings and dual F-theory with unbroken $\mathfrak{e}_7\oplus\mathfrak{e}_7$ algebra was constructed by the authors in \cite{Clingher2018}. 
\par Constructions of six-dimensional (6D) non-geometric heterotic strings with unbroken $\mathfrak{e}_8\oplus\mathfrak{e}_7$ gauge algebra and dual F-theory were also discussed in \cite{MM}. Related to the extremal fibrations discussed in section \ref{sec3.1}, 6D F-theory models on elliptically fibered CY 3-folds, constructed by fibering extremal K3 fibrations over $\mathbb{P}^1$, were analyzed in \cite{Kimura1810, Kimura201902}.

\section*{Acknowledgments}

We would like to thank Shigeru Mukai for discussions.

\end{document}